\documentclass[aip,reprint]{revtex4-1}

\usepackage{graphicx}

\usepackage{amssymb}
\usepackage{amsmath} 

\begin{document}


\title{Adversarial control of synchronization in complex oscillator networks} 



\author{Yasutoshi Nagahama}
\affiliation{ 
Department of Bioscience and Bioinformatics, Kyushu Institute of Technology, Iizuka, Fukuoka, Japan
}%
\author{Kosuke Miyazato}
\affiliation{ 
Department of Bioscience and Bioinformatics, Kyushu Institute of Technology, Iizuka, Fukuoka, Japan
}%
\author{Kazuhiro Takemoto}%
\email{takemoto@bio.kyutech.ac.jp.}
\affiliation{
Department of Bioscience and Bioinformatics, Kyushu Institute of Technology, Iizuka, Fukuoka, Japan
}%
\affiliation{
Data Science and AI Research Center, Kyushu Institute of Technology, Iizuka, Fukuoka, Japan
}%


\date{\today}

\begin{abstract}
This study investigates perturbation strategies inspired by adversarial attack principles from deep learning, designed to control synchronization dynamics through strategically crafted weak perturbations. We propose a gradient-based optimization method that identifies small phase perturbations to dramatically enhance or suppress collective synchronization in Kuramoto oscillator networks. Our approach formulates synchronization control as an optimization problem, computing gradients of the order parameter with respect to oscillator phases to determine optimal perturbation directions. Results demonstrate that extremely small phase perturbations applied to network oscillators can achieve significant synchronization control across diverse network architectures. Our analysis reveals that synchronization enhancement is achievable across various network sizes, while synchronization suppression becomes particularly effective in larger networks, with effectiveness scaling favorably with network size. The method is systematically validated on canonical model networks including scale-free and small-world topologies, and real-world networks representing power grids and brain connectivity patterns. This adversarial framework represents a novel paradigm for synchronization management by introducing deep learning concepts to networked dynamical systems.
\end{abstract}


\maketitle 

\textbf{
Synchronization emerges naturally in many networked systems, where individual components coordinate their behavior to act collectively. Controlling such synchronization in complex networks has been challenging, typically requiring substantial interventions. This study introduces a novel approach inspired by adversarial attack principles from artificial intelligence, where tiny, strategically designed perturbations can dramatically alter system behavior. We applied this concept to networks of oscillating systems and discovered that extremely small adjustments can either enhance or suppress synchronization across the entire network. These weak interventions become particularly effective in larger networks, suggesting excellent scalability for real-world applications. We validated this approach on various network types including power grids and brain networks, demonstrating potential applications for infrastructure stability and managing pathological brain activity. This research offers a new paradigm for controlling complex networked systems with weak intervention.
}

\section{Introduction}
Synchronization phenomena in complex networks of coupled oscillators are ubiquitous in nature, spanning from the coordinated flashing of fireflies and cardiac pacemaker cells to the stability of power grids and brain network dynamics \cite{arenas2008synchronization,suykens2008introduction,dorfler2014synchronization}.
The Kuramoto model has emerged as a fundamental paradigm for understanding these collective behaviors by focusing on the phase dynamics of weakly coupled oscillators \cite{acebron2005kuramoto,rodrigues2016kuramoto}.
Despite its apparent simplicity, the model captures essential features of synchronization transitions and has become an invaluable theoretical framework for investigating synchronization control strategies across diverse scientific disciplines.

Traditional synchronization control approaches in Kuramoto oscillator networks primarily target the system's intrinsic parameters. Network topology significantly influences synchronization dynamics, with scale-free networks promoting synchronization through highly connected hub nodes \cite{moreno2004synchronization,peron2019onset,medvedev2020kuramoto} and small-world networks enhancing synchronization via shortcuts that reduce effective path lengths between distant oscillators \cite{hong2002synchronization,grabow2010small,nishikawa2003heterogeneity}. Based on this understanding, topology modification approaches strategically add or remove connections or implement adaptive rewiring to alter critical coupling thresholds \cite{zhou2006dynamical,chen2008synchronization,zhang2014synchronization,pinto2015optimal,papadopoulos2017development,ghorban2022rewiring,ruangkriengsin2023low,song2024network}, though such structural modifications require physical access to network connections in biological or infrastructure systems. Natural frequency manipulation exploits the principle that narrow frequency distributions promote synchronization \cite{acebron1998adaptive,taylor2010spontaneous, gil2021optimally}, while maintaining non-natural frequencies requires external energy sources. Furthermore, coupling strength adjustment represents another approach \cite{ren2007adaptive,ha2016synchronization,cho2023interplay}, with specific connection weight distributions such as frequency-dependent coupling influencing synchronization \cite{wang2011synchronization}, though this can lead to complex dynamical behaviors \cite{rossi2023shifts}. Optimization approaches for natural frequencies or link weights have also been proposed \cite{menara2022functional}, with computational requirements scaling with network size.

External intervention strategies control synchronization without permanently altering system characteristics. Pinning control applies targeted signals to strategically selected nodes to drive network-wide synchronization \cite{wang2002pinning,sorrentino2007controllability,xiang2007pinning,yang2010pinning,lin2016controlling}, with performance depending on appropriate node selection in heterogeneous networks. Time-delayed feedback and global feedback mechanisms (such as mean-field feedback) can promote or suppress synchronization by monitoring the collective state of the system and adjusting control signals accordingly \cite{yeung1999time,rosenblum2004delayed,taher2019enhancing,franci2012desynchronization,ozawa2021feedback}, requiring optimization of delay times and feedback strengths, with performance being sensitive to these parameter settings. Noise-induced synchronization exploits the principle that appropriately tuned random perturbations can enhance collective behavior through mechanisms such as stochastic resonance and stochastic synchronization \cite{kawamura2007noise,esfahani2012noise,lai2013noise,tyloo2019noise}, though the stochastic nature of this approach introduces inherent limitations for deterministic control applications. The sensitivity of networked oscillators to strategic perturbations has been further demonstrated in studies of network vulnerabilities, where targeted interventions at individual nodes can produce significant changes in global synchronization behavior \cite{tyloo2023assessing}.

While these approaches have made valuable contributions, the above challenges remain. To address these limitations, we explore an alternative strategy inspired by adversarial attack principles from deep learning. Adversarial attacks involve strategically crafted small perturbations added to inputs that can dramatically alter the outputs of deep neural network models, exploiting inherent system vulnerabilities \cite{yuan2019adversarial,hirano2020simple,hirano2020vulnerability,hirano2021universal,chakraborty2021survey,akhtar2021advances,koga2022simple}. This concept has recently been extended to dynamics on networks in other domains, specifically applied to voter model dynamics and prisoner's dilemma games on networks, where small strategic perturbations can significantly alter collective behaviors \cite{chiyomaru2022adversarial,chiyomaru2023mitigation,mizutaka2023crossover,takemoto2024steering,ninomiya2025mitigating}. We introduce this adversarial framework to synchronization control in Kuramoto oscillator networks, where the goal is to identify small perturbations to the networked dynamics that can disrupt or enhance synchronization by exploiting the system's intrinsic sensitivities. This approach offers unique characteristics: requiring only weak perturbations to phases while leveraging optimization principles to achieve large-scale effects on collective behavior. By treating synchronization control as an adversarial problem, we introduce a novel theoretical framework that complements existing approaches and provides new insights into the fundamental mechanisms of synchronization manipulation. While our method draws inspiration from adversarial attacks in deep learning, specifically the principle of using small perturbations to achieve large effects, it is fundamentally a gradient-based control strategy that requires real-time access to system states rather than a machine learning technique. We emphasize that our approach represents a conceptual transfer of the adversarial principle rather than establishing direct mathematical analogies between neural networks and oscillator networks.

In this work, we develop and analyze adversarial attack strategies specifically designed for Kuramoto oscillator networks to achieve synchronization control through phase perturbations. We formulate the adversarial attack as an optimization problem that identifies near-minimal perturbations to oscillator phases that can maximally disrupt or enhance network synchronization. Our approach combines gradient-based optimization techniques with network analysis to systematically exploit the vulnerabilities inherent in the oscillator networks. We demonstrate that synchronization can indeed be effectively controlled through this adversarial framework via extensive numerical simulations on both model networks and real-world networks. Moreover, we investigate and discuss the effects of network size, average node degree, and network connectivity patterns on the outcomes of adversarial attacks.

\section{Model}
We consider the Kuramoto model on networks \cite{moreno2004synchronization,arenas2008synchronization,dorfler2014synchronization,rodrigues2016kuramoto,peron2019onset}, where the time evolution of the phases $\theta_i$ ($i=1,\dots,N$) of $N$ coupled oscillators on a given network is described by
\begin{equation}
    \frac{\mathrm{d}}{\mathrm{d}t}\theta_i = \omega_i + K\sum_{j=1}^NA_{ij}\sin(\theta_j - \theta_i),
\end{equation}
where $\omega_i$ represents the natural frequency of oscillator $i$, and $K$ denotes the coupling strength that characterizes the strength of interactions between oscillators.
For simplicity, we consider unweighted and undirected networks.
Thus, $A_{ij} = A_{ji} = 1$ if there exists a link between oscillators (nodes) $i$ and $j$, and $A_{ij} = 0$ otherwise.

Alternative formulations using degree normalized coupling strength $K/k_i$ \cite{arenas2008synchronization,rodrigues2016kuramoto}, where $k_i$ is the degree of node $i$, do not affect our main conclusions, as will be shown when we develop our gradient based attack methodology.

To quantify the degree of synchronization in the network, we define the complex order parameter
\begin{equation}
    Re^{i\psi}=\frac{1}{N}\sum_{j=1}^Ne^{i\theta_j},
\end{equation}
where $R$ represents the synchronization strength and $\psi$ denotes the average phase. The synchronization strength $R$ ranges from 0 to 1, with $R \approx 1$ indicating strong synchronization (all oscillators have similar phases) and $R \approx 0$ indicating an incoherent state (oscillator phases are uniformly distributed).

\section{Adversarial attcks}
Here, we apply the concept of adversarial attacks to promote or suppress synchronization in Kuramoto oscillator networks. Specifically, we use the order parameter $R$ as the objective function and apply a gradient descent method to adjust each oscillator's phase along the gradient direction, thereby guiding the entire system toward a synchronized state.

The gradient-based approach is formulated as
\begin{equation}
    \theta_i \leftarrow \theta_i + \epsilon \frac{\partial R}{\partial \theta_i},
\end{equation}
where the gradient $\partial R/\partial \theta_i$ is derived as
\begin{equation}
    \frac{\partial R}{\partial \theta_i} = \frac{1}{N}\sin(\psi - \theta_i).
\end{equation}
Here, $\psi$ represents the average phase calculated just before applying the perturbation.
Physically, this gradient indicates the direction that brings each oscillator's phase closer to the average phase, thereby promoting synchronization.

However, obtaining exact phase values and controlling perturbation strength can be challenging. To avoid these limitations, following the principle of the fast gradient sign method \cite{goodfellow2015explaining}, we consider using only the sign of the gradient. Specifically, we add perturbations to each oscillator's phase at regular time intervals $\tau$ as follows:
\begin{equation}
    \theta_i \leftarrow \theta_i + \epsilon \times \mathrm{sign}\left[\sin(\psi -\theta_i)\right],
\end{equation}

When $\epsilon < 0$, the adjustment occurs in the opposite direction of the gradient, actively suppressing synchronization.

The parameter $\epsilon$ controls the perturbation strength, where $|\epsilon|$ determines the magnitude of phase perturbations applied to the network.

To compare the performance of the adversarial perturbations with random controls, we also consider random perturbations with the same perturbation strength applied at the same time intervals $\tau$: $\theta_i \leftarrow \theta_i + \epsilon \times s$, where $s$ is a random variable uniformly sampled from the set $\{-1, +1\}$.

\section{Simulations}
\subsection{Network models and simulation setup}
Our numerical analysis examines the effectiveness of adversarial attacks on synchronization dynamics using three well-established network topologies that capture different structural properties found in real-world systems.

We employ networks consisting of $N = 1000$ oscillators with an average degree $\langle k \rangle = 6$, unless stated otherwise. Robustness tests confirm that varying network size ($N \in [200, 4000]$) or average degree ($\langle k \rangle \in [4, 20])$ does not alter the qualitative behavior. Each oscillator's intrinsic frequency $\omega_i$ follows a standard normal distribution $N(0,1)$, while initial phases are randomly assigned from a uniform distribution over $[0, 2\pi]$.

Three network architectures are investigated: Erd\H{o}s--R\'enyi (ER) random networks \cite{albert2002statistical,Takemoto2012_book} serve as our baseline topology, where edges are placed between $L$ randomly chosen node pairs, yielding Poisson-distributed degrees with mean $\langle k \rangle= 2L/N$.
To capture heterogeneous connectivity patterns, we also examine Barab\'asi--Albert (BA) scale-free networks \cite{albert2002statistical,barabasi1999emergence} constructed through preferential attachment, where new nodes connect to $m$ existing nodes, producing power-law degree distributions $P(k) \propto k^{-3}$ with $\langle k \rangle= 2m$.
Additionally, Watts--Strogatz (WS) small-world networks \cite{watts1998collective} are analyzed, which begin as regular lattices where each node connects to $k$ nearest neighbors, then edges are rewired with probability of 0.05 \cite{hirano2019difficulty,chiyomaru2022adversarial,takemoto2024steering}, creating networks with enhanced clustering compared to random graphs.

The dynamics are integrated until the system reaches equilibrium, and the order parameter $R$ is measured in this steady state.
The time evolution of $R$ is computed for each realization, and statistical reliability is ensured by averaging results across 100 independent trials, each with different network topologies and initial phase configurations.

\subsection{Adversarial perturbation effects on synchronization transitions}

\begin{figure}[htbp]
\includegraphics[width=70mm]{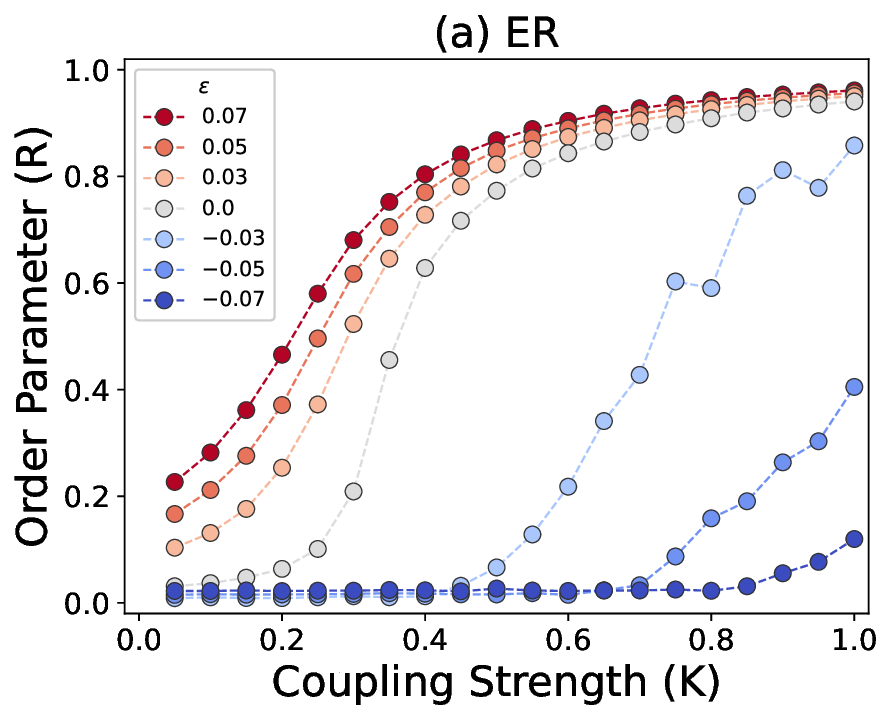} \\
\includegraphics[width=70mm]{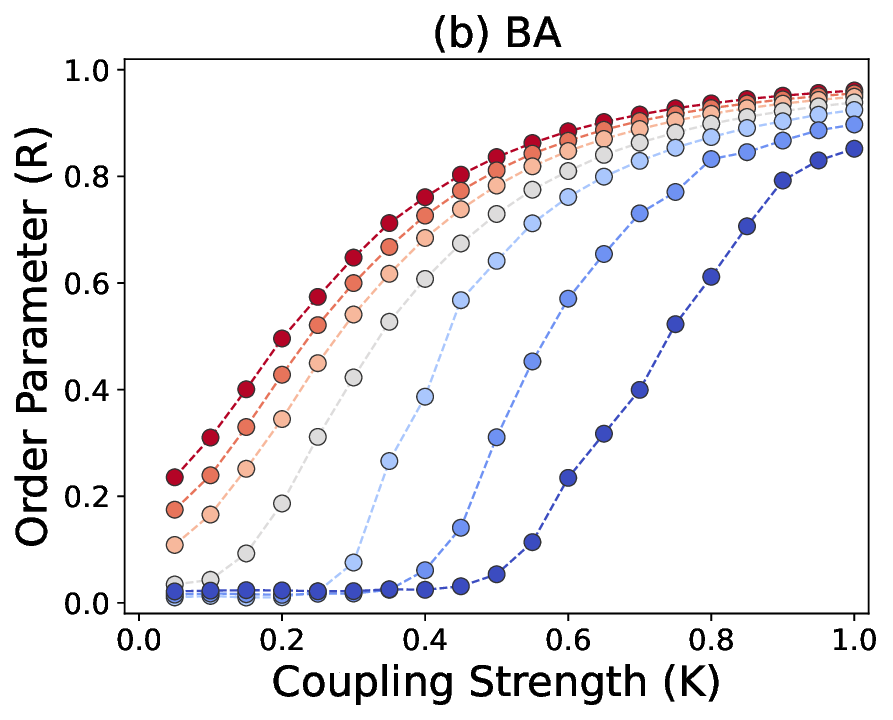} \\
\includegraphics[width=70mm]{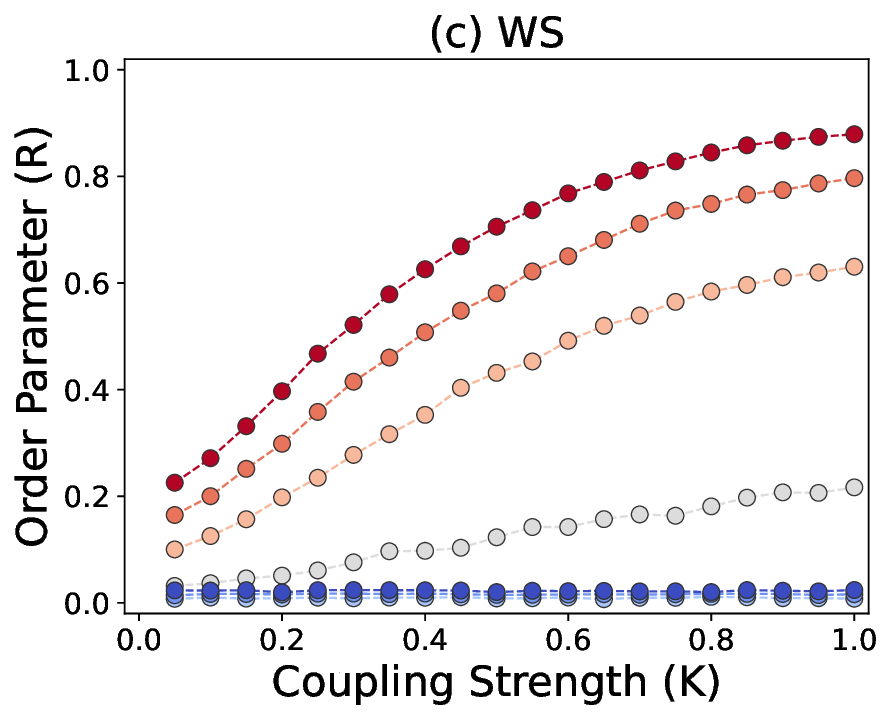}
\caption{\label{fig:R_vs_K_model} Synchronization transitions under adversarial perturbations. Order parameter $R$ versus coupling strength $K$ for (a) Erd\H{o}s–R\'enyi (ER), (b) Barab\'asi--Albert (BA), and (c) Watts--Strogatz (WS) networks with different perturbation parameters $\epsilon$. Positive $\epsilon$ (red) promotes synchronization, negative $\epsilon$ (blue) suppresses it. Intervention interval $\tau = 0.3$.}%
\end{figure}

Figure \ref{fig:R_vs_K_model} shows the relationship between the order parameter $R$ and coupling strength $K$ for different perturbation parameters $\epsilon$ across three network topologies. The most striking observation is that even small perturbation strengths can significantly alter synchronization transitions regardless of network topology. Positive $\epsilon$ values (red curves) consistently promote synchronization by shifting the transition to lower coupling strengths, while negative $\epsilon$ values (blue curves) suppress synchronization, requiring higher coupling strengths to achieve the same level of synchronization across all network types. This demonstrates the remarkable effectiveness of adversarial perturbations in controlling synchronization dynamics with weak intervention strength.

While the fundamental ability to manipulate synchronization transitions is preserved across different network architectures, some topology-dependent variations are observed. BA networks (Fig. \ref{fig:R_vs_K_model}b) show somewhat reduced susceptibility to synchronization suppression compared to ER networks (Fig. \ref{fig:R_vs_K_model}a), likely due to their scale-free structure with highly connected hubs that facilitate synchronization. Conversely, WS networks (Fig. \ref{fig:R_vs_K_model}c) exhibit particularly pronounced responses to positive perturbations.
Although WS networks with their lattice-like local structure are inherently difficult to synchronize, adversarial interventions can dramatically enhance their synchronization capability.

\subsection{Perturbation strength effects}
\begin{figure*}[htbp]
\includegraphics[width=50mm]{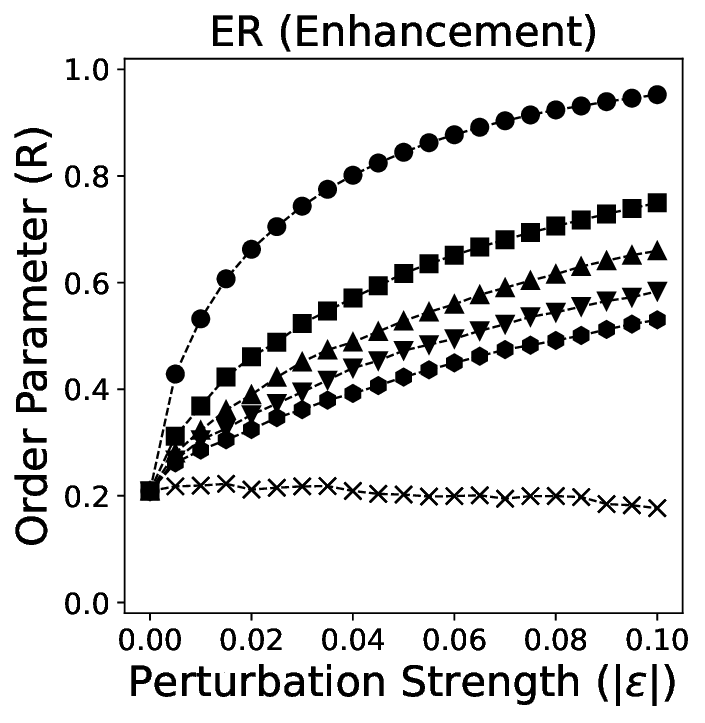}
\includegraphics[width=50mm]{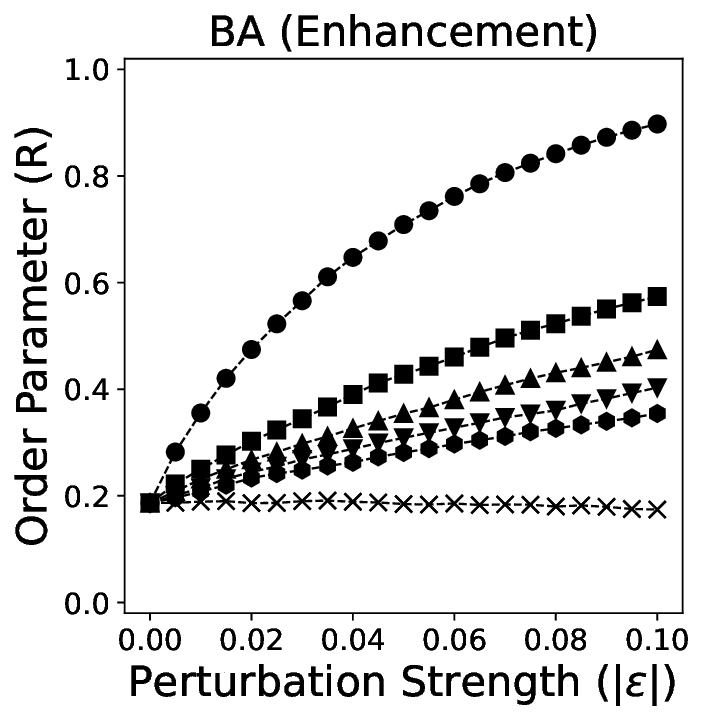} 
\includegraphics[width=50mm]{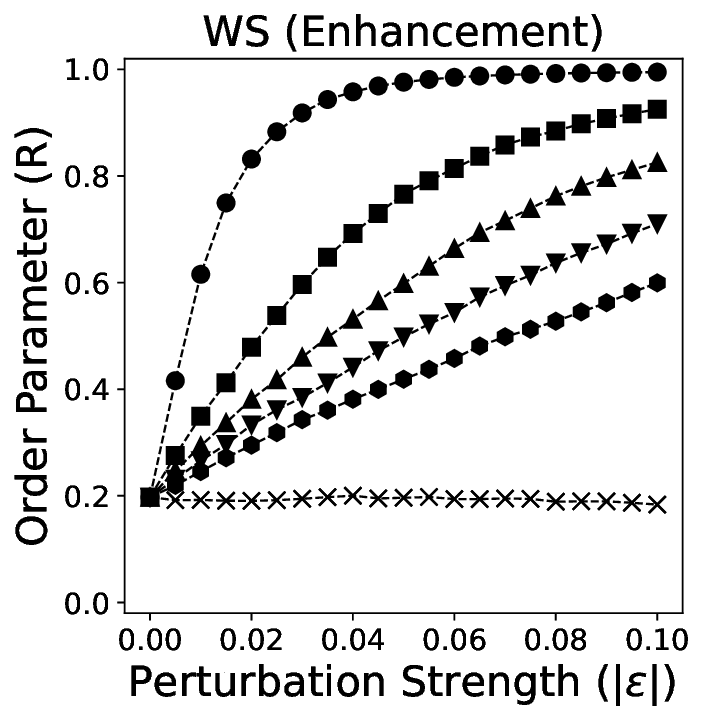} \\
\includegraphics[width=50mm]{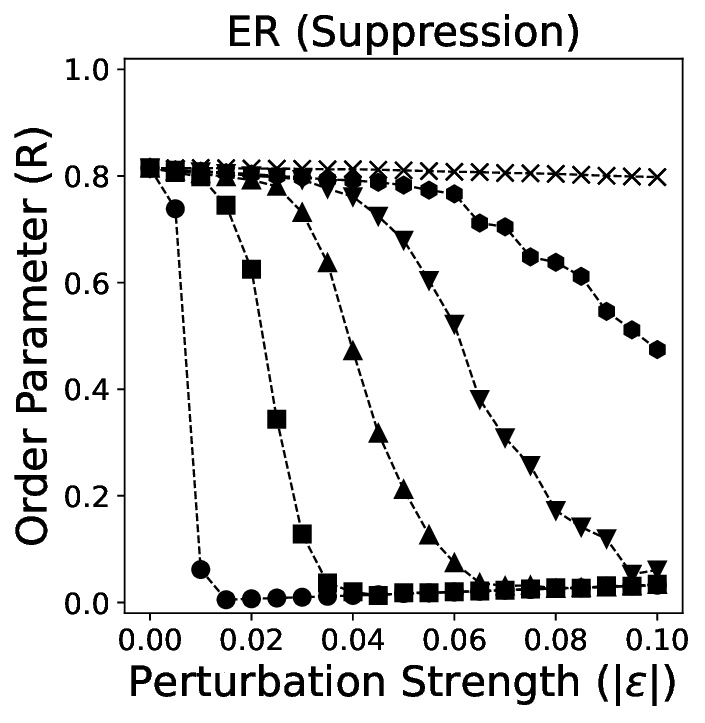} 
\includegraphics[width=50mm]{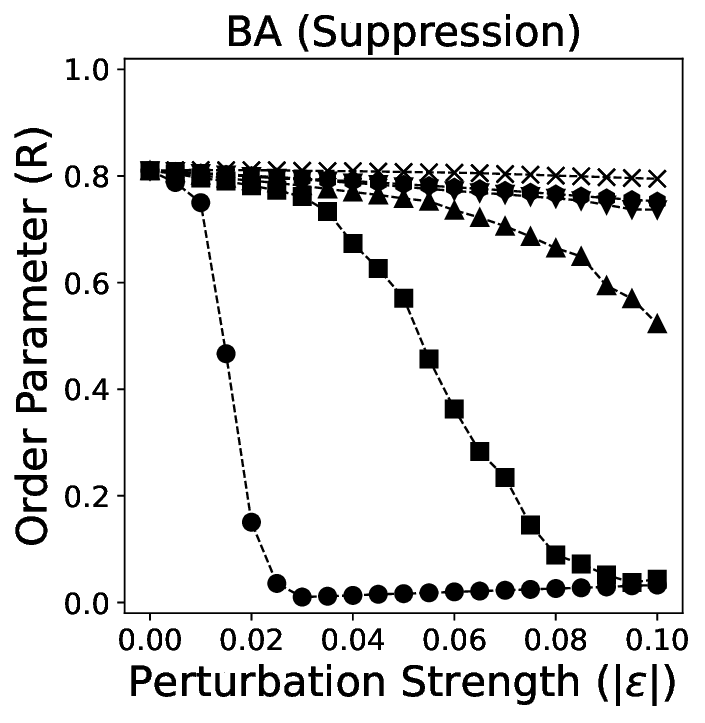}
\includegraphics[width=50mm]{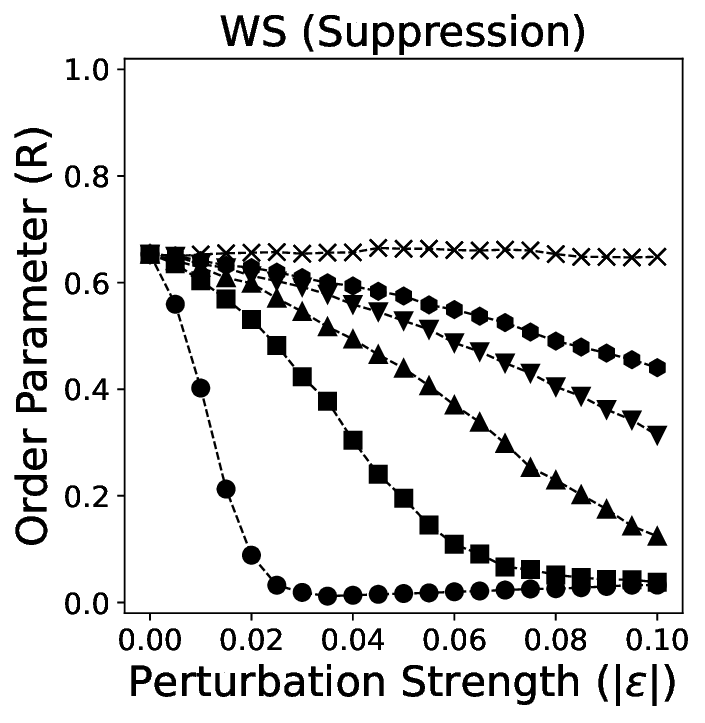} 
\caption{\label{fig:R_vs_eps} Order parameter $R$ as a function of perturbation strength $|\epsilon|$ for (left to right) Erd\H{o}s--R\'enyi (ER), Barab\'asi--Albert (BA), and Watts--Strogatz (WS) networks. Upper panels: enhancement from $R \approx 0.2$ ($K = 0.2$, 0.3, 0.85 for ER, BA, WS respectively); lower panels: suppression from $R \approx 0.8$ ($K = 0.55$, 0.6, 6.0). Filled symbols indicate adversarial attack intervention intervals for a: {\Large $\bullet$} ($\tau = 0.1$), $\blacksquare$ ($\tau = 0.3$), $\blacktriangle$ ($\tau = 0.5$), $\blacktriangledown$ ($\tau = 0.7$), and 
hexagons ($\tau = 1.0$). Crosses shows random perturbation controls for $\tau = 0.1$.}%
\end{figure*}

To investigate the relationship between perturbation strength and synchronization control more systematically, Figure \ref{fig:R_vs_eps} examines the order parameter $R$ as a function of perturbation strength $|\epsilon|$ for fixed coupling strength across different network topologies and intervention intervals $\tau$. The coupling strength $K$ is chosen to ensure appropriate baseline conditions: for synchronization enhancement (upper panels), we select $K$ values that yield $R \approx 0.2$ in the unperturbed state, while for synchronization suppression (lower panels), we choose $K$ values corresponding to $R \approx 0.8$. However, due to the inherent difficulty of synchronization in WS networks, we empirically set $K=6.0$ for this topology to achieve meaningful perturbation effects.

The results demonstrate a striking sensitivity of synchronization dynamics to perturbation strength. For synchronization enhancement (upper panels), $R$ increases sharply with even small positive $\epsilon$ values, showing that weak perturbations can yield substantial synchronization promotion. Similarly, for synchronization suppression (lower panels), $R$ decreases rapidly as the magnitude of negative $\epsilon$ increases, indicating that small perturbation strengths are sufficient to significantly disrupt collective dynamics. This sharp dependence on $\epsilon$ underscores the remarkable efficiency of adversarial attacks in manipulating synchronization with weak intervention.

The intervention interval $\tau$ influences the steepness of these transitions. Shorter intervals ($\tau = 0.1$) produce the sharpest $R$--$\epsilon$ curves, while longer intervals ($\tau = 1.0$) result in more gradual changes. This reflects the expected reduction in control effectiveness with less frequent interventions.

The comparison with random perturbations (crosses) reveals the superiority of adversarial attacks. Random perturbations show minimal variation in $R$ regardless of perturbation strength, remaining close to the unperturbed baseline. This stark contrast highlights that the effectiveness of adversarial perturbations stems from their strategic targeting rather than mere noise addition.

Network-dependent variations are also evident. WS networks exhibit the strongest responses to both enhancement and suppression attempts, while BA networks show the most resistance to perturbation effects, particularly for synchronization suppression.

\subsection{Network size effects}

\begin{figure*}[htbp]
\includegraphics[width=50mm]{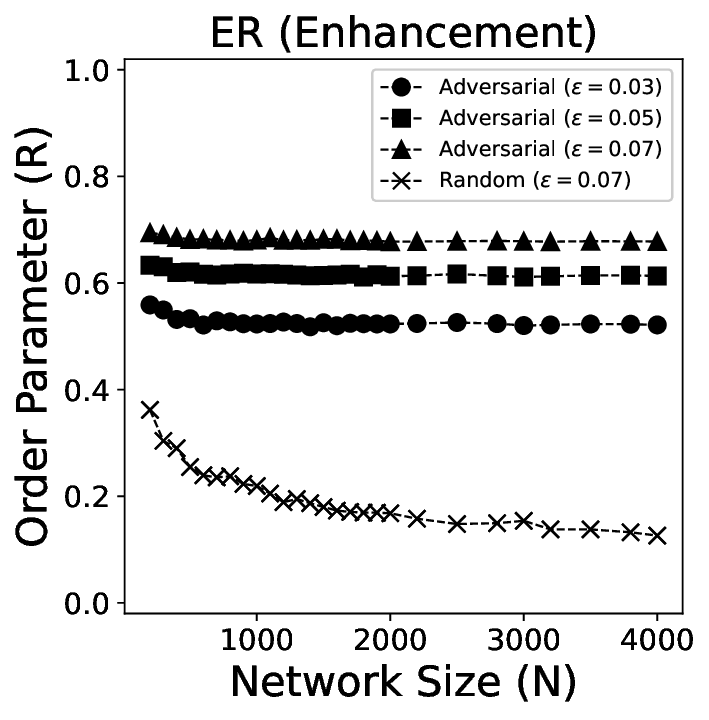}
\includegraphics[width=50mm]{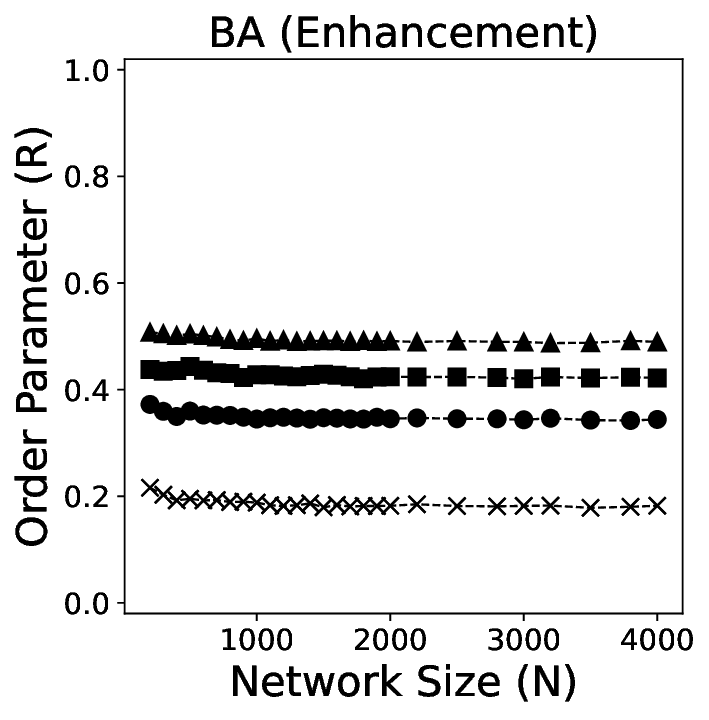} 
\includegraphics[width=50mm]{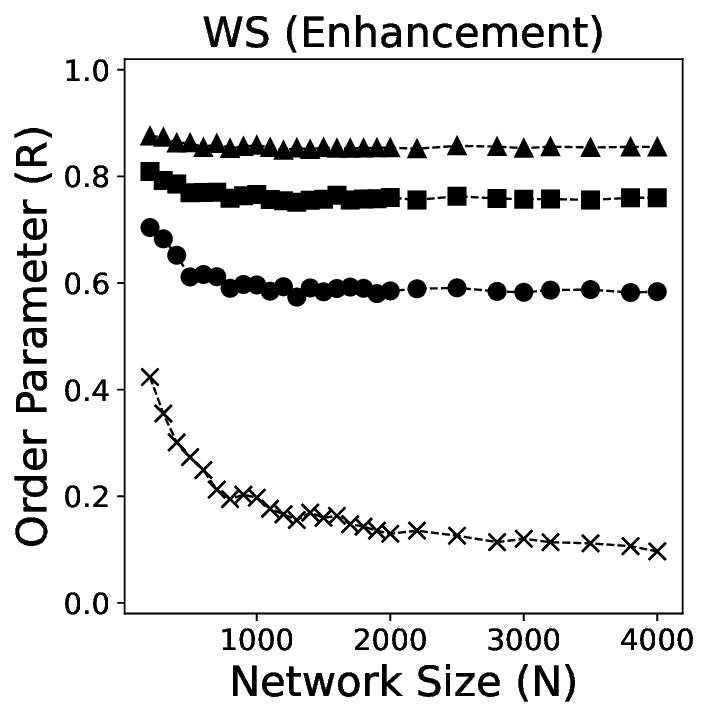} \\
\includegraphics[width=50mm]{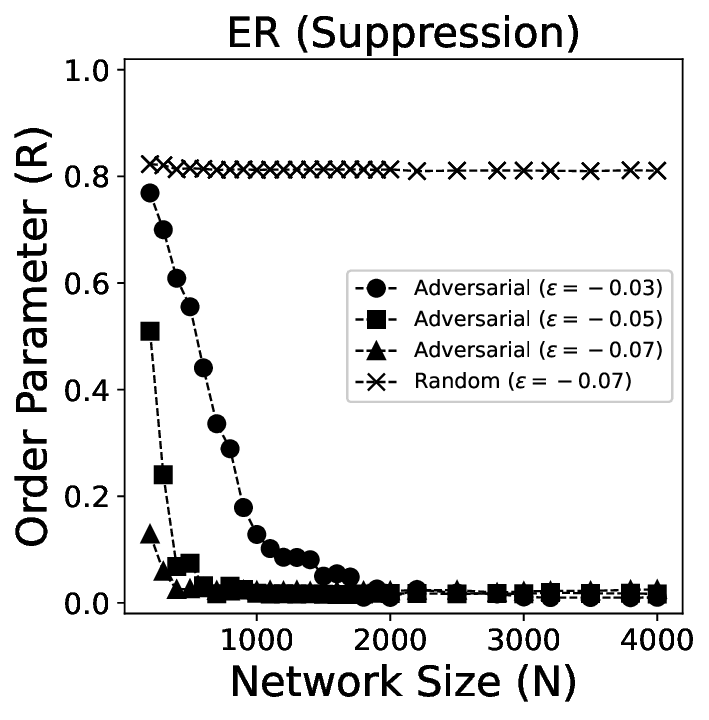} 
\includegraphics[width=50mm]{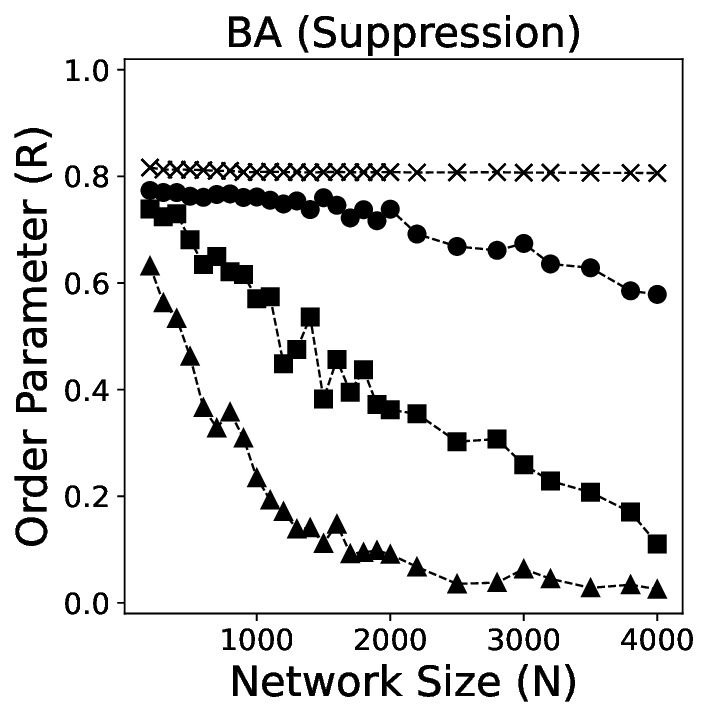}
\includegraphics[width=50mm]{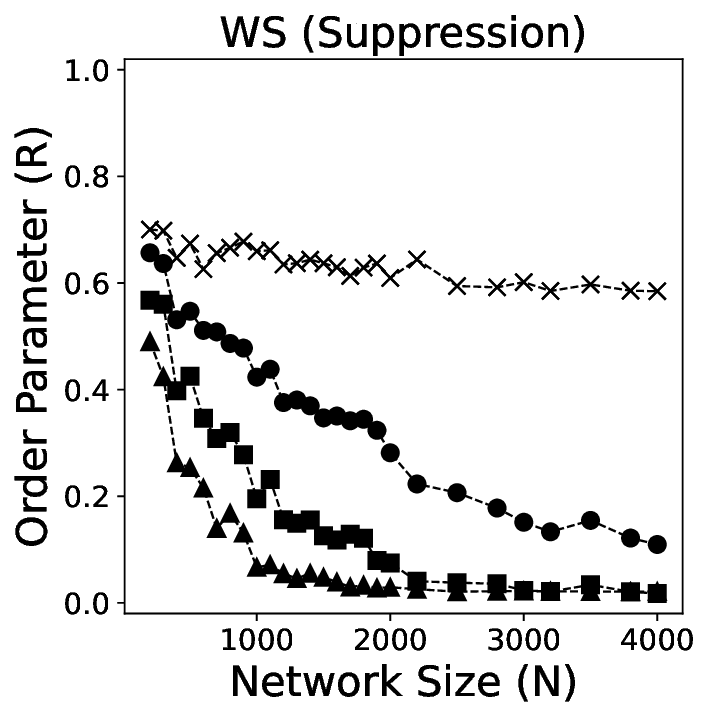}
\caption{\label{fig:R_vs_N} Order parameter $R$ as a function of network size $N$ for (left to right) Erd\H{o}s--R\'enyi (ER), Barab\'asi--Albert (BA), and Watts--Strogatz (WS) networks. Intervention interval $\tau = 0.3$. Upper panels: enhancement from $R \approx 0.2$ ($K = 0.2$, 0.3, 0.85 for ER, BA, WS respectively); lower panels: suppression from $R \approx 0.8$ ($K = 0.55$, 0.6, 6.0). Filled symbols indicate different perturbation parameters $\epsilon$. Crosses show random perturbation controls.}%
\end{figure*}

To examine how network size influences the effectiveness of adversarial attacks, Figure \ref{fig:R_vs_N} shows the order parameter $R$ as a function of network size $N$ using the same coupling strength $K$ settings as Fig. \ref{fig:R_vs_eps} but with the intervention interval fixed at $\tau = 0.3$. The results reveal distinctly different behaviors between synchronization enhancement and suppression scenarios.

For synchronization enhancement (upper panels), the effectiveness of adversarial attacks shows minimal dependence on network size across all three topologies. The order parameter remains relatively constant as $N$ increases from 500 to 4000, indicating that positive perturbations maintain their synchronization-promoting capabilities regardless of system scale. This size-independence suggests that the mechanisms underlying synchronization enhancement are robust to network dimensionality.

In stark contrast, synchronization suppression (lower panels) exhibits a pronounced network size dependence across all network topologies. The effectiveness of adversarial attacks increases substantially with larger networks, as evidenced by the progressive decrease in $R$ with increasing $N$. This trend is observed in all three network types, though BA networks show the most resistance to suppression due to their inherently higher synchronizability. The enhanced suppression effectiveness in larger networks stems from the inherent difficulty of achieving synchronization in high-dimensional systems, where the increased degrees of freedom naturally oppose collective dynamics.

Crucially, the comparison with random perturbations (crosses) confirms that this size-dependent enhancement of suppression is genuinely due to adversarial targeting rather than general noise effects. Random perturbations show little variation with network size and remain close to baseline values, demonstrating that the observed trends are specific to the strategic nature of adversarial attacks. This finding highlights a fundamental asymmetry: while larger networks become increasingly vulnerable to adversarial desynchronization, they do not become more susceptible to random disruption, underscoring the sophisticated targeting capability of adversarial perturbations.

\subsection{Real-world network applications}
\begin{figure}[htbp]
\includegraphics[width=70mm]{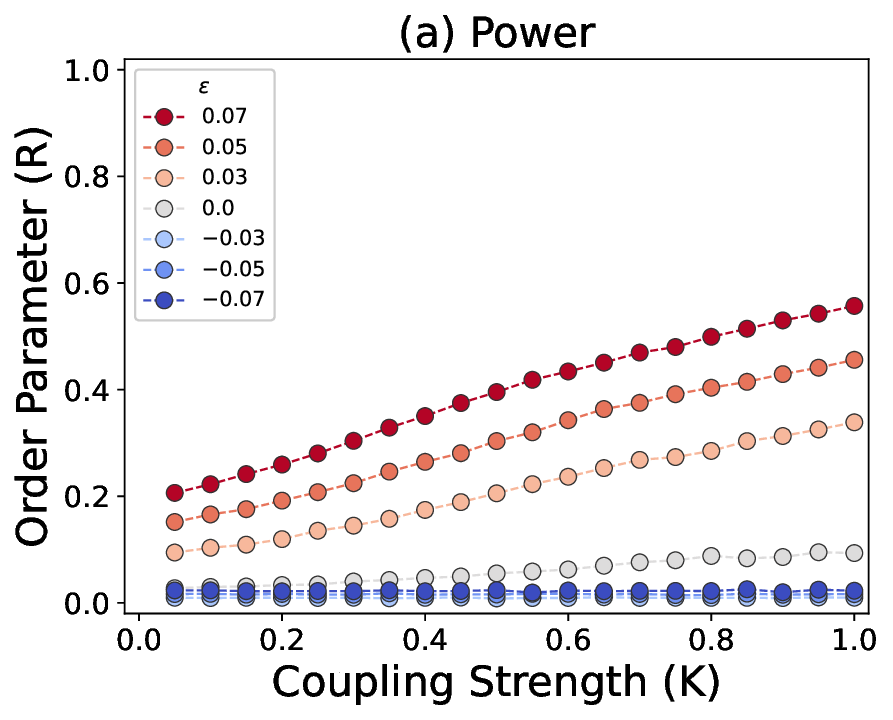} \\
\includegraphics[width=70mm]{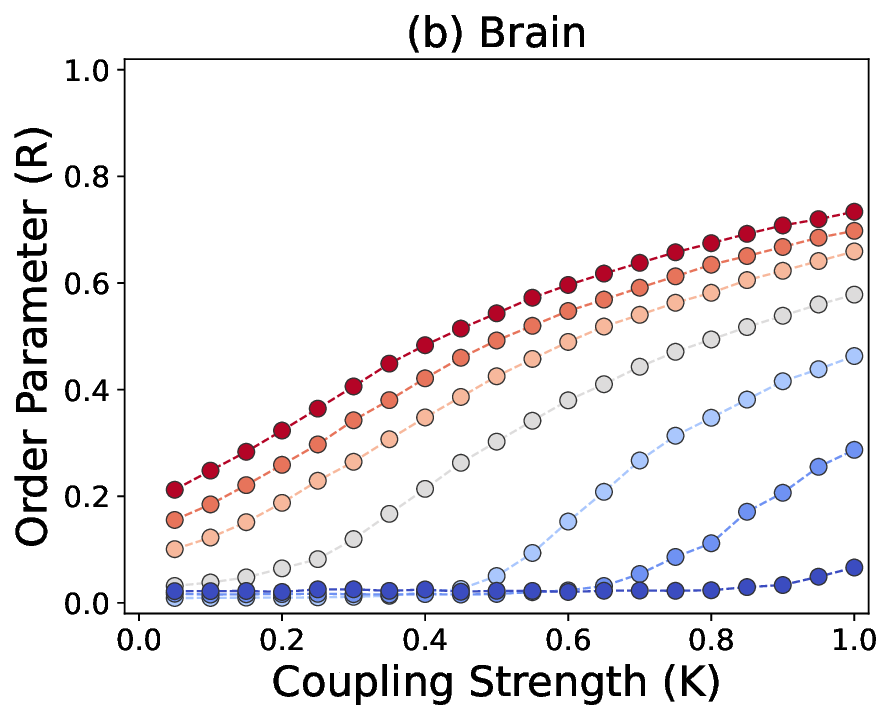}
\caption{\label{fig:R_vs_K_real-world} Order parameter $R$ versus coupling strength $K$ for (a) power network ($N=1138$ and $\langle k \rangle=2.6$) and (b) brain network ($N=987$ and $\langle k \rangle=3.1$) with different perturbation parameters $\epsilon$.  Positive $\epsilon$ (red) promotes synchronization, negative $\epsilon$ (blue) suppresses it. Intervention interval $\tau = 0.3$.}%
\end{figure}

To demonstrate the practical applicability of our adversarial attack methodology, we applied it to two real-world networks where synchronization plays a crucial role: a power network (1138-bus system \cite{networkdatarep}) and a mouse brain network \cite{amunts2013bigbrain,networkdatarep}. For simplicity, both networks were treated as unweighted undirected networks, and only the largest connected component was used for analysis. Figure \ref{fig:R_vs_K_real-world} shows the order parameter $R$ as a function of coupling strength $K$ for both networks under various perturbation conditions.

Both networks exhibit behavior consistent with our findings from model networks, confirming the generalizability of adversarial attacks on network synchronization. In the power network (Fig. \ref{fig:R_vs_K_real-world}a), which shows inherently low synchronization due to its sparse structure, adversarial perturbations demonstrate particularly effective synchronization enhancement. Positive perturbations successfully elevate the network's collective behavior across the entire coupling range, with stronger perturbations producing substantial improvements in coordination. This enhancement capability is especially valuable for power grids, where improved frequency synchronization directly translates to system stability \cite{carrasco2006power,li2017optimizing}.

The brain network (Fig. \ref{fig:R_vs_K_real-world}b) demonstrates highly effective responses to both enhancement and suppression attacks. Positive perturbations produce strong synchronization enhancement, substantially elevating the order parameter across all coupling values. Conversely, the network shows significant vulnerability to desynchronization attacks, with negative perturbations effectively maintaining low synchronization levels even under strong coupling conditions that would normally induce robust collective behavior. This suppression capability is particularly relevant for therapeutic applications \cite{tass2002desynchronization,chazottes2005desynchronization,tass2007therapeutic,andrzejak2016all}, as pathological hypersynchronization in brain networks is associated with epileptic seizures \cite{lehnertz2009synchronization,jiruska2013synchronization}.

These results demonstrate that adversarial attacks on network synchronization represent a genuine concern for real-world systems, while simultaneously suggesting potential applications for controlled synchronization enhancement in cases where increased coordination is desired.

\section{Discussion}
In this study, we have demonstrated that adversarial attack strategies, originally developed for deep learning systems, can be effectively adapted to control synchronization in networked dynamical systems governed by the Kuramoto model. Our results consistently show that small, strategically applied perturbations to the phases of network nodes can dramatically enhance or suppress collective synchronization across diverse network architectures, from theoretical frameworks to practical infrastructure and biological systems.

Our results confirm that borrowing the adversarial attack concept from the deep learning domain and applying it to network synchronization control represents a significant advance.
Unlike traditional synchronization control methods that typically require structural modifications \cite{zhou2006dynamical,chen2008synchronization,zhang2014synchronization,pinto2015optimal,papadopoulos2017development,ghorban2022rewiring,ruangkriengsin2023low,song2024network,menara2022functional} such as adding or removing links, rewiring connections, or introducing external driving forces \cite{wang2002pinning,sorrentino2007controllability,xiang2007pinning,yang2010pinning,lin2016controlling,yeung1999time,rosenblum2004delayed,taher2019enhancing,franci2012desynchronization,ozawa2021feedback,kawamura2007noise,esfahani2012noise,lai2013noise,tyloo2019noise}, our adversarial perturbation strategy achieves effective control through weak interventions. While this approach builds on well-established principles of gradient-based feedback and weak intervention strategies, our contribution lies in the specific formulation and application context that leverages the order parameter structure for distributed phase perturbations. While conventional approaches often necessitate global knowledge of network topology or significant alterations to the system architecture, our method operates by applying small amplitude perturbations ($|\epsilon| \leq 0.1$) to the phases of network nodes, requiring only the computation of gradients and maintaining the original network structure intact.

It is important to distinguish our approach from related work on network perturbations, particularly Byzantine attacks on oscillator networks \cite{tyloo2023assessing}. While both approaches demonstrate the significant impact that strategic perturbations can have on networked oscillators, they differ fundamentally in purpose and methodology. Byzantine attacks involve malicious actors who completely control individual nodes with arbitrary input signals to disrupt or destabilize synchronization for adversarial purposes. In contrast, our gradient-based framework uses the mathematical structure of the order parameter to determine optimal perturbation directions that enhance or suppress synchronization for beneficial control objectives. Rather than arbitrary signal injection, our method employs principled optimization to achieve desired synchronization states while maintaining the integrity of the network structure. This represents a constructive application of strategic perturbation principles, transforming concepts originally associated with network vulnerabilities into tools for beneficial synchronization management.

This efficiency is particularly notable when compared to conventional control approaches. Many existing synchronization control strategies involve node selective interventions \cite{wang2002pinning,sorrentino2007controllability,xiang2007pinning,yang2010pinning,lin2016controlling}, structural network modifications \cite{zhou2006dynamical,chen2008synchronization,zhang2014synchronization,pinto2015optimal,papadopoulos2017development,ghorban2022rewiring,ruangkriengsin2023low,song2024network}, or the application of external driving forces \cite{yeung1999time,rosenblum2004delayed,taher2019enhancing,franci2012desynchronization,ozawa2021feedback,kawamura2007noise,esfahani2012noise,lai2013noise,tyloo2019noise} to specific system components. In contrast, our adversarial approach demonstrates that modest perturbations applied uniformly across the network can produce system wide changes in synchronization behavior. The distributed nature of this intervention offers complementary advantages for applications where uniform access to network nodes is feasible, providing an alternative pathway for synchronization control that maintains the original network architecture.

Moreover, our findings reveal the dual nature of network vulnerability and controllability: the same mechanisms that make networks susceptible to adversarial attacks also provide powerful tools for beneficial synchronization control. This insight opens new avenues for both understanding network robustness and developing efficient control strategies for complex systems where synchronized behavior is either desired or needs to be prevented.

Although we focused on the Kuramoto model in this study, the core of our attack methodology, namely the gradient computation from the order parameter, is derived purely from the mathematical definition of synchronization and is not specific to any particular oscillator model. This generality suggests that our adversarial perturbation approach can be readily extended to other oscillatory systems, such as FitzHugh--Nagumo neurons \cite{gerster2020fitzhugh,al2024criticality}, Stuart--Landau oscillators \cite{choe2010controlling,selivanov2012adaptive}, or any dynamical system where a meaningful order parameter can be defined. The mathematical framework we developed is thus broadly applicable to diverse classes of networked dynamical systems.

While our results demonstrate the effectiveness of adversarial perturbations for synchronization control, we acknowledge important practical considerations for real-world implementation. The current approach requires simultaneous access to all oscillator phases and the ability to perturb all nodes, which presents challenges in practical systems such as power grids or brain networks. We view our current results as establishing a theoretical foundation that opens promising avenues for future developments.

Building on this foundation, several exciting research directions emerge: developing sparse perturbation schemes that achieve effective control by targeting only strategically selected nodes, creating partial-access strategies that operate with limited system information, and designing adaptive methods that can function under realistic constraints. These developments represent natural extensions of our framework that could enable practical synchronization control in real-world networked systems while maintaining the core advantages of weak intervention and distributed control.

Additionally, the optimal balance between perturbation parameter $\epsilon$ and intervention timing $\tau$ requires careful consideration. In this study, $\tau$ values were selected empirically based on the trade-off between control effectiveness and intervention frequency, as demonstrated in our parameter sweep analysis (Fig. \ref{fig:R_vs_eps}). While larger perturbations and more frequent interventions naturally enhance control effectiveness, practical applications demand minimal interference with the system. The key challenge is achieving robust synchronization control using the smallest possible perturbations applied as infrequently as possible. A deeper theoretical analysis of the attack mechanism would provide valuable insights into these parameter trade-offs and the fundamental principles underlying the observed synchronization control.

Future work should also validate our approach across different oscillator models, as well as extend the framework to directed and weighted networks \cite{porfiri2008synchronization,duan2008synchronization,tiberi2017synchronization}, and systems with higher-order interactions \cite{gao2023dynamics,liu2024synchronization,wang2024synchronization}. These extensions would further demonstrate the broad applicability of our adversarial synchronization control paradigm.

Nonetheless, the core finding that small, strategically applied perturbations can dramatically alter collective synchronization behavior represents a significant conceptual advance. The simplicity of the approach, combined with its effectiveness across diverse network architectures, establishes a new paradigm for synchronization control that bridges deep learning and network dynamics. The broad scope for future developments, from sparse interventions to higher-order networks, underscores the fundamental importance and potential impact of this research direction.

The data and code used in this study are publicly available at \url{https://github.com/kztakemoto/advSync}.


%
%

%

\begin{acknowledgments}
This study was funded by JSPS KAKENHI (grant number JP21H03545).
\end{acknowledgments}

\bibliography{references}

\end{document}